\documentclass{cimento1}
\usepackage{amsmath,graphicx}

\title{The hadronic vacuum polarization contribution to $(g_\mu - 2)$:\\ Lattice QCD+QED calculations \thanks{{\it Presented by Davide Giusti at ``XVIII edition of Incontri di Fisica delle Alte Energie'', Napoli (Italy), 08--10 April 2019.}}}
\shorttitle{The HVP contribution to $(g_\mu - 2)$: Lattice QCD+QED calculations}
\author{D.~Giusti\from{ins:x}\from{ins:y} \atque S.~Simula\from{ins:y}}
\instlist{
\inst{ins:x}
Dipartimento di Matematica e Fisica, Universit\`a degli Studi Roma Tre - Roma, Italy
\inst{ins:y}
INFN, Sezione di Roma Tre - Roma, Italy
}

\begin{document}

\maketitle

\begin{abstract}
The anomalous magnetic moment of the muon $a_\mu$ is one of the most accurate quantities in Particle Physics.
The long-standing discrepancy of about $3.7$ standard deviations between the experimental value and the prediction of the Standard Model could represent an intriguing indication of New Physics.
The experiments at Fermilab (E989) and at J-PARC (E34) aim at reducing significantly the experimental uncertainty, thus making the theoretical one due to hadronic corrections the main limitation of this stringent test of the Standard Model.
In this contribution we present the results of a first-principles lattice calculation of the hadronic vacuum polarization (HVP) contribution to $a_\mu$, including electromagnetic and $SU(2)$-breaking corrections.
Our determination, $a_\mu^{\rm HVP} = 682.0 ~ (18.7) \cdot 10^{-10}$, turns out to be in agreement with recent theoretical determinations based on the dispersive analyses of the experimental cross section data for the annihilation process $e^+e^- \to hadrons$.
Furthermore, we provide for the first time a lattice estimate for the missing part of $a_\mu^{\rm HVP}$ not covered in the MUonE experiment, $\left[a_\mu^{\rm HVP}\right]_> = 91.6 ~ (2.0) \cdot 10^{-10}$.
\end{abstract}

\section{Introduction}

The anomalous magnetic moment of the muon, $a_\mu = (g_\mu - 2) / 2$, is one of the most precisely-determined observables in Particle Physics, having been measured with an accuracy of $0.54 ~ \mbox{ppm}$ by experiment E821 at Brookhaven National Laboratory (BNL)~\cite{ref:Bennett:2006fi}.
Since the anomaly is mediated by quantum-mechanical loops, the muon $(g_\mu - 2)$ could provide one of the most valuable probes of new degrees of freedom beyond the Standard Model (SM).
The present SM theory value lies below the BNL E821 measurement by more than three standard deviations~\cite{ref:Tanabashi:2018oca}.
To assess whether this intriguing deviation is due to new particles or interactions, both the theory and measurement uncertainties must be improved.
In this respect, experiment E989 at Fermi National Accelerator Laboratory, running since March 2018, aims at reducing the total error on $a_\mu$ achieved by E821 by a factor of four~\cite{ref:Logashenko:2015xab}.
A second, complementary experiment (E34), planned to begin taking data around 2020, is being mounted at J-PARC~\cite{ref:Otani:2015lra}.
In parallel, numerous efforts are underway by the theoretical community to tackle the hadronic loop contributions due to the HVP and hadronic light-by-light terms (an updated status of the theoretical efforts can be found in ref.~\cite{ref:gm2_Th}), which are the largest sources of error in the SM result.
As for the dominant HVP contribution, the most precise determinations, quoting relative errors of about $0.4 \div 0.6 \,\%$, are obtained by combining experimental measurements of electron-positron inclusive scattering into hadrons with dispersion relations (see ref.~\cite{Davier:2019can} and therein quoted).
Here we present the results of a completely independent cross-check study of $a_\mu^{\rm HVP}$ performed from first principles using the latest numerical lattice QCD+QED simulations.

\section{Lattice QCD+QED calculations of $a_\mu^{\rm HVP}$: results}

In this section we summarize the main lattice QCD+QED results we have obtained in refs.~\cite{ref:Giusti:2017jof,ref:Giusti:2018mdh,ref:Giusti:2019xct}, using the gauge configurations generated by the ETM Collaboration with $N_f = 2+1+1$ dynamical quarks, at three values of the lattice spacing $a$ varying from $0.089$ to $0.062 ~ \mbox{fm}$, at several lattice sizes ($L \simeq 1.8 \div 3.5 ~ \mbox{fm}$) and with pion masses in the range $M_\pi \simeq 220 \div 490 ~ \mbox{MeV}$.

After the extrapolation of our lattice data to the physical pion mass and to the continuum and infinite-volume limits, we get the following determinations for the ${\cal O}(\alpha_{e.m.}^2)$ light-, strange- and charm-quark-connected HVP contributions~\cite{ref:Giusti:2017jof,ref:Giusti:2018mdh}
 \begin{eqnarray}
      a_\mu^{\rm HVP} (ud) & = & 619.0 ~ (17.8) \cdot 10^{-10} ~ , \label{e.amu_LO_ud}\\
      a_\mu^{\rm HVP} (s) & = & 53.1 ~ (2.5) \cdot 10^{-10} ~ , \label{e.amu_LO_s}\\
      a_\mu^{\rm HVP} (c) & = & 14.75 ~ (0.56) \cdot 10^{-10} ~ . \label{e.amu_LO_c}
\end{eqnarray}

Furthermore, we have performed a lattice calculation of the leading-order electromagnetic (e.m.) and strong isospin-breaking (IB) corrections contributing to the HVP to ${\cal O}(\alpha_{e.m.}^3)$ and ${\cal O}(\alpha_{e.m.}^2 (m_d - m_u) / \Lambda_{\rm QCD})$, respectively.
We find~\cite{ref:Giusti:2019xct}
\begin{equation}
      \delta a_\mu^{\rm HVP} (udsc) = 7.1 ~ (2.9) \cdot 10^{-10} ~ ,
      \label{e.amu_IB}
\end{equation}
which represents the most accurate determination of the IB corrections to $a_\mu^{\rm HVP}$ to date.

Adding the determinations of eqs.~(\ref{e.amu_LO_ud})-(\ref{e.amu_IB}) and an estimate of the quark-disconnected diagrams, $a_\mu^{\rm HVP} (disconn.) = -12 ~ (4) \cdot 10^{-10}$, obtained using the findings of refs.~\cite{ref:Borsanyi:2017zdw,ref:Blum:2018mom}, we finally get for the muon $a_\mu^{\rm HVP}$ the value
\begin{equation}
      a_\mu^{\rm HVP} = 682.0 ~ (18.7) \cdot 10^{-10} ~ ,
      \label{e.amu_tot}
\end{equation}
which agrees within the errors with the recent determinations based on dispersive analyses of the experimental cross section data for $e^+e^-$ annihilation into hadrons~\cite{Davier:2019can}.

Recently~\cite{ref:Calame:2015fva} it has been proposed to determine $a_\mu^{\rm HVP}$ by measuring the running of $\alpha_{e.m.}(q^2)$ for space-like values of the squared four-momentum transfer $q^2$ using a muon beam on a fixed electron target. 
The method is based on the following alternative formula for calculating $a_\mu^{\rm HVP}$~\cite{ref:Lautrup:1971jf}:
\begin{equation}
    a_\mu^{\rm HVP} = \frac{\alpha_{e.m.}}{\pi} \int_0^1 dx \, (1 - x) \, \Delta \alpha_{e.m.}^{\rm HVP}[q^2(x)] ~ ,
    \label{eq:amu_muon}
\end{equation}
where $\Delta \alpha_{e.m.}^{\rm HVP}(q^2)$ is the hadronic contribution to the running of $\alpha_{e.m.}(q^2)$ evaluated at $q^2(x) \equiv - m_\mu^2 \, x^2 / (1-x)$.
The quantity $\Delta \alpha_{e.m.}^{\rm HVP}(q^2)$ can be extracted from the $q^2$-dependence of the $\mu e \to \mu e$ cross section data after the subtraction of the leptonic and weak contributions~\cite{ref:Calame:2015fva}.
For the proposed MUonE experiment exploiting the muon beam at the CERN North Area~\cite{ref:MUonE} the region $x \in [0.93, 1]$ in eq.~(\ref{eq:amu_muon}) cannot be reached and, therefore, the corresponding contribution, hereafter indicated by $\left[a_\mu^{\rm HVP}\right]_>$, needs to be estimated using either $e^+ e^-$ data or lattice QCD simulations.

In ref.~\cite{ref:Giusti:2018mdh} we have provided the first lattice estimate of the lowest-order light-quark-connected contribution to $\left[a_\mu^{\rm HVP}\right]_>$, which is found to be equal to
\begin{equation}
    \left[a_\mu^{\rm HVP}\right]_> (ud) = 81.2 ~ (1.7) \cdot 10^{-10} ~ .
    \label{e.amu_MUonE_ud}
\end{equation}

\begin{figure}[htb!]
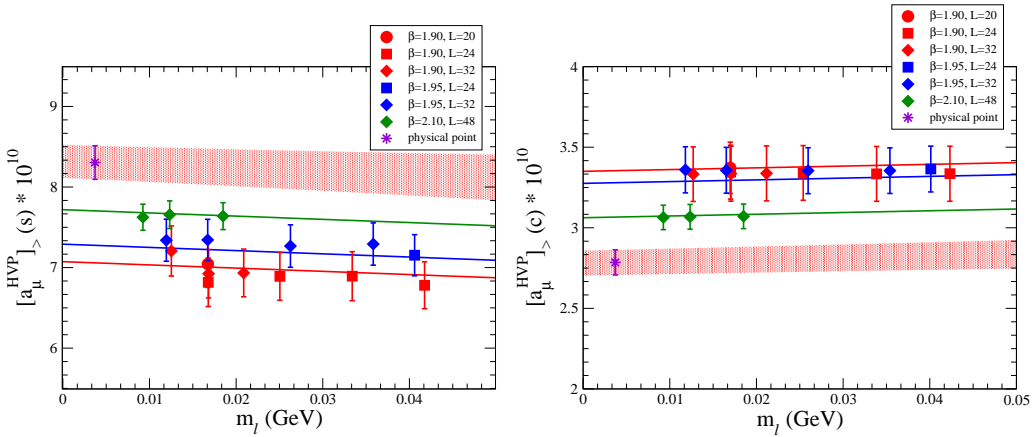

  \begin{center}
    \includegraphics[scale=0.33]{strange}~~\includegraphics[scale=0.33]{charm}
  \end{center}
\caption{Results for the leading-order strange- (left panel) and charm-quark (right panel) contributions to $\left[a_\mu^{\rm HVP}\right]_>$ versus the renormalized light-quark mass $m_\ell$ in the $\overline{\rm MS} (2 ~ \mbox{GeV})$ scheme. The solid lines correspond to the results of the combined fits (\ref{e.fit_amu_MUonE}) obtained in the infinite-volume limit at each value of the lattice spacing. The purple asterisks represent the results of $\left[a_\mu^{\rm HVP}\right]_> (s)$ and $\left[a_\mu^{\rm HVP}\right]_> (c)$ extrapolated to the physical pion mass, corresponding to $m_\ell^{phys} = 3.70 ~ (0.17) ~ \mbox{MeV}$ (determined in ref.~\cite{ref:Carrasco:2014cwa}), and to the continuum and infinite-volume limits, while the red areas indicate the corresponding uncertainties at the level of one standard deviation. Errors are statistical only.}
\label{fig.amu_MUonE_sc}
\end{figure}

Here we present new determinations for the strange- and charm-quark terms as well as for the IB effects.
In fig.~\ref{fig.amu_MUonE_sc} we show the results for $\left[a_\mu^{\rm HVP}\right]_> (s)$ (left panel) and $\left[a_\mu^{\rm HVP}\right]_> (c)$ (right panel) from the individual gauge ensembles (see ref.~\cite{ref:Giusti:2017jof} for further details).

We perform combined fits for the extrapolations to the physical pion mass, the continuum and infinite-volume limits using the same fitting functions (4.2) of ref.~\cite{ref:Giusti:2017jof}, namely
\begin{equation}
    \left[a_\mu^{\rm HVP}\right]_> (s,c) = A_0^{s,c} \left[ 1 + A_1^{s,c} \, \xi + D^{s,c} \, a^2 + F^{s,c} \, \xi \, \frac{e^{-M_\pi L}}{M_\pi L} \right] ~ ,
    \label{e.fit_amu_MUonE}
\end{equation}
where $\xi \equiv M_\pi^2 / (4 \pi f_0)^2$, $f_0$ is one of the QCD low-energy constants to leading order and $A_0^{s,c}, A_1^{s,c}, D^{s,c}, F^{s,c}$ are free parameters determined by the fitting procedures.
We consider either a constant $(A_1^{s,c} = 0)$ or a linear $(A_1^{s,c} \neq 0)$ chiral dependence of the lattice data in order to estimate the systematic uncertainties due to the chiral extrapolations to the physical pion mass.
Moreover, the exponential term in eq.~(\ref{e.fit_amu_MUonE}) is a phenomenological representation of finite-volume corrections.
For our ${\cal O} (a)$-improved simulation setup, discretization effects are parameterized by the term proportional to $a^2$.
The results of the linear fits (\ref{e.fit_amu_MUonE}) are shown in fig.~\ref{fig.amu_MUonE_sc} by the solid lines.

At the physical point we obtain
 \begin{eqnarray}
      \left[a_\mu^{\rm HVP}\right]_> (s) & = & 8.3 ~ (0.4) \cdot 10^{-10} ~ , \label{e.amu_MUonE_s}\\
      \left[a_\mu^{\rm HVP}\right]_> (c) & = & 2.8 ~ (0.1) \cdot 10^{-10} ~ . \label{e.amu_MUonE_c}
\end{eqnarray}
The uncertainties of the above findings represent the sum in quadrature of various sources of errors, namely statistical, fitting procedure, input parameters, discretization, scale setting, finite-volume effects and chiral extrapolation.
The error budget is obtained as described in ref.~\cite{ref:Giusti:2017jof}.

In addition, we here provide the first lattice estimates of the leading-order e.m.~and strong IB corrections to the light-, strange- and charm-quark (connected) contributions to $\left[a_\mu^{\rm HVP}\right]_>$.
The lattice calculation is performed within the RM123 approach~\cite{ref:deDivitiis:2011eh,ref:deDivitiis:2013xla}, which consists in the expansion of the path integral in powers of the $u$- and $d$-quark mass difference $(m_d - m_u)$ and of the e.m.~coupling $\alpha_{e.m.}$, and in the quenched-QED approximation, {\it i.e.}~treating dynamical quarks as electrically neutral particles.
Each of the three contributions is evaluated following the same strategy adopted for the corresponding term in the case of the muon~\cite{ref:Giusti:2019xct}.
Recent non-perturbative determinations of the QED corrections to the relevant renormalization constants are included (see table V of ref.~\cite{ref:Giusti:2019xct}).
The extrapolations to the physical pion mass and the continuum and infinite-volume limits are performed using the same fitting functions employed in refs.~\cite{ref:Giusti:2019xct,ref:Giusti:2019dmu}.

Without reporting the details of the above procedures, we here limit ourselves to give the results of the lattice calculations at the physical point.
Our determinations are
\begin{eqnarray}
      \left[\delta a_\mu^{\rm HVP}\right]_> (ud) & = & 0.9 ~ (0.3) \cdot 10^{-10} ~ , \label{e.amu_MUonE_IB_ud}\\
      \left[\delta a_\mu^{\rm HVP}\right]_> (s) & = & -0.0005 ~ (0.0004) \cdot 10^{-10} ~ , \label{e.amu_MUonE_IB_s}\\
      \left[\delta a_\mu^{\rm HVP}\right]_> (c) & = & 0.0034 ~ (0.0007) \cdot 10^{-10} ~ . \label{e.amu_MUonE_IB_c}
\end{eqnarray}
Adding the above findings we get
\begin{equation}
      \left[\delta a_\mu^{\rm HVP}\right]_> (udsc) = 0.9 ~ (0.3) \cdot 10^{-10} ~ ,
      \label{e.amu_MUonE_IB}
\end{equation}
whose uncertainty includes also an estimate of the error due to the quenched-QED approximation according to ref.~\cite{ref:Giusti:2019xct}.

As for the estimate of the quark-disconnected contribution we adopt the following strategy.
We first consider the ratio of quark-disconnected over quark-connected (the sum of the determinations in eqs.~(\ref{e.amu_LO_ud})-(\ref{e.amu_LO_c})) contributions in the case of the muon, namely $-12 \, (4) \, / \, 686.9 \, (18.0) = 0.0175 ~ (59)$.
Then, the same value of the ratio is assumed to hold as well for $\left[a_\mu^{\rm HVP}\right]_>$.
This implies $\left[a_\mu^{\rm HVP}\right]_> (disconn.) = -1.6 ~ (0.5) \cdot 10^{-10}$.
We choose to double the uncertainty, thus adopting the following conservative estimate
\begin{equation}
      \left[a_\mu^{\rm HVP}\right]_> (disconn.) = -1.6 ~ (1.0) \cdot 10^{-10} ~ .
      \label{e.amu_MUonE_IB_disc}
\end{equation}

Adding all the various contributions (\ref{e.amu_MUonE_ud}),\,(\ref{e.amu_MUonE_s}),\,(\ref{e.amu_MUonE_c}),\,(\ref{e.amu_MUonE_IB}),\,(\ref{e.amu_MUonE_IB_disc}), we finally obtain the following value for the missing part of $a_\mu^{\rm HVP}$ not covered in the MUonE experiment
\begin{equation}
      \left[a_\mu^{\rm HVP}\right]_> = 91.6 ~ (2.0) \cdot 10^{-10} ~ .
      \label{e.amu_MUonE}
\end{equation}
The uncertainty of our determination is close to statistical error expected in the MUonE experiment for the complementary contribution $\left[a_\mu^{\rm HVP}\right]_< \equiv \left[a_\mu^{\rm HVP}\right] - \left[a_\mu^{\rm HVP}\right]_>$ after two years of data taking at the CERN North Area.

\acknowledgments
D.G.~warmly thanks the organizers for the very stimulating conference held in Naples.

\end{document}